\documentstyle[twoside,fleqn,espcrc2]{article}

\input epsf.sty
\newdimen\psfigsize
\def\psfigure#1 #2 #3 #4 #5{
    \begin{figure}[tbh]
    \vbox{
    \null\vskip-0.2in\hskip#2\epsfxsize=#1 \epsfbox[0 0 4096 4096]{#4}
    \vskip -0.3in
    \caption {#5 \label{#3}}
    \vskip -0.20truein plus0.10truein}
    \end{figure}
}
\def\psoddfigure#1 #2 #3 #4 #5 #6{
    \begin{figure}[tbh]
    \vbox{
    \null\vskip-0.2in\hskip#3\epsfxsize=#1 \epsfbox[0 0 4096 4096]{#5}
    \vskip -#1 \vskip #2 \vskip 10truept
    \vskip -0.2in
    \caption {#6 \label{#4}}
    \vskip 0.0truein plus0.2truein}
    \end{figure}
}
\def\BE{\begin{equation}}
\def\EE{\end{equation}}
\def\BEA{\begin{eqnarray}}
\def\EEA{\end{eqnarray}}

\newcommand{\AmS}{{\protect\the\textfont2
  A\kern-.1667em\lower.5ex\hbox{M}\kern-.125emS}}

\title{Recent MILC spectrum results
\thanks{presented by R.L.~Sugar}}

\author{ C.~Bernard
\address{Department of Physics, Washington University, St.~Louis, MO 63130, USA}% "a"
, T. Blum
\address{Department of Physics, Brookhaven National Lab, Upton, NY 11973, USA} %"b"
, T.~A.~DeGrand
\address{Physics Department, University of Colorado, Boulder, CO 80309, USA} % "c"
, C.~DeTar
\address{Physics Department, University of Utah, Salt Lake City, UT 84112, USA} % "d"
, Steven~Gottlieb
\address{Department of Physics, Indiana University, Bloomington, IN 47405, USA} % "e"
, Urs~M.~Heller
\address{SCRI, Florida State University, Tallahassee, FL 32306-4052, USA} %"f"
, J.~Hetrick
\address{Department of Physics, University of Arizona, Tucson, AZ 85721, USA} % "g"
, C.~McNeile$\,\null^{\rm d}$
, K.~Rummukainen$\,\null^{\rm e}$
, R.L.~Sugar
\address{Department of Physics, University of California, Santa Barbara, CA 93106, USA}
, Doug~Toussaint$\,\null^{\rm g}$
and M.~Wingate$\,\null^{\rm c}$
} %end \author
\begin{document}

\begin{abstract}
We report on results from three spectrum calculations with staggered
quarks: 1) a quenched calculation with the standard action for the gluons 
and quarks; 2) a quenched calculation with improved actions for both the gluons
and quarks; and 3) a calculation with two flavors of dynamical quarks
using the standard actions for the gluons and quarks.
\end{abstract}
\maketitle

\section{INTRODUCTION}

A controlled calculation of the masses of the light hadrons has
been one of the major goals of lattice QCD\cite{STEVEG}. It has become 
increasingly clear in recent years that in order to carry out such
a calculation one must perform high statistics simulations using
a range of lattice volumes in order to control finite size effects,
a range of quark masses in order to extrapolate to the chiral limit,
and a range of lattice spacings in order to extrapolate to the
continuum limit. We have carried out such
a series of simulations in the quenched approximation with staggered
quarks\cite{STEVELAT95}. In Sec.~2 we report on new results
from this study with particular emphasis on the extrapolation to
the chiral limit.

Improved actions have the potential to significantly 
increase our ability to perform accurate spectrum calculations\cite{IMPROVED}.
In Sec.~3 we report on exploratory quenched calculations 
using improved actions for the gluons and the staggered
quarks. Finally, in Sec.~4 we show preliminary results
from a series of full QCD calculations with two flavors of
staggered quarks at gauge coupling $6/g^2=5.50$.

\section{QUENCHED SPECTRUM WITH STAGGERED QUARKS}

Over the past several years we have carried out a series of quenched
spectrum calculations with staggered quarks. Calculations were
performed with four values of the gauge coupling, $6/g^2=5.54$, 5.70,
5.85, and 6.15. Five different lattice volumes were used at 5.70 and three
at 5.85 in order to study finite volume effects. Five quark masses
were used at each coupling. For $6/g^2=5.70$ and 5.85, we used
$am_q=0.16$, 0.08, 0.04, 0.02 and 0.01. At 6.15 the masses were halved,
and at 5.54 they were doubled.
For each coupling, lattice volume and quark mass we performed
correlated fits to determine the masses. 
An Edinburgh plot is shown in Fig.~1.  

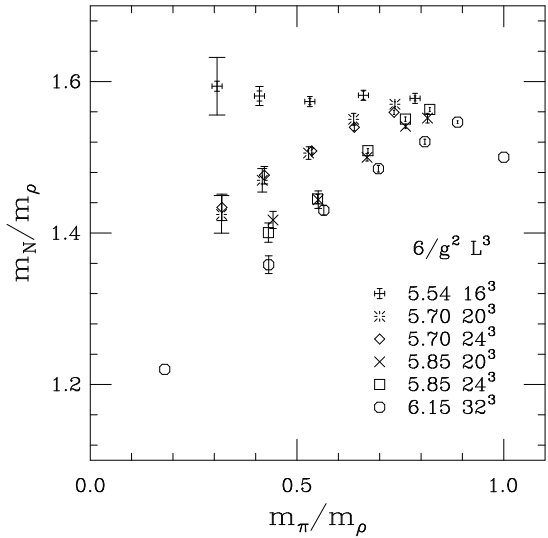
\psfigure 3.0in -0.15in {FIG1} {fig1.ps} {Edinburgh plot for the
quenched staggered spectrum.}

Hadron masses calculated on the same set of gauge configurations,
but with different quark masses are, of course, correlated. So,
we performed correlated fits for each coupling and lattice volume
to extrapolate to the chiral limit. We used a wide variety of
fitting functions with terms inspired by quenched chiral perturbation
theory. The first conclusion was that simple linear or quadratic
fits to the nucleon and rho masses are inconsistent with our
data. Good fits are possible. 
For example, the function $M+am_q + bm_q^{{3/2}}+cm_q^2$
gives reasonable fits for both the nucleon and rho masses. 
We also obtain good fits for the nucleon with functions containing
square roots or logarithms of the quark masses suggested by
quenched chiral perturbation theory.
Once such a fit is in hand, we can investigate the dependence 
of $m_N/m_\rho$ as a function of lattice spacing for fixed $m_\pi/m_\rho$. 
Some typical plots are shown in Fig.~2 using the above fitting form. 
These fits allow us to extrapolate to the continuum limit for
fixed values of $m_\pi/m_\rho$. They are also proving very useful in
comparing results obtained with improved actions to our high statistics
results with the standard staggered action. 

\psfigure 3.0in -0.15in {FIG2} {fig2.ps} {$m_N/m_\rho$ as a function
of lattice spacing for fixed $m_\pi/m_\rho$.} 

In Fig.~3 we plot $m_N/m_\rho$ as a function of $am_\rho(m_q=0)$ 
with the ratio $m_\pi/m_\rho$ held fixed at its physical value.
Fits to the rho and nucleon used in this plot were again made
with the form $M+am_q + bm_q^{{3/2}}+cm_q^2$.
The solid curve is a fit of the form $c_0+c_2a^2$ to data
for all four couplings, and the dotted line is a fit of
the same form to data for the three weakest couplings.
Here $a$ is the lattice spacing, and $c_i$ are fitted parameters.
We are unable to uniquely determine
the optimum fitting function for the chiral extrapolation from
our data. In particular, nucleon fits with $m_q^{{1/2}}$ terms 
have confidence levels similar to those shown in Figs.~2 and 3.
An $m_q^{{1/2}}$ terms tend to pull the value of $m_N$
down in the chiral limit, thereby reducing $m_N/m_\rho$. 
At present the uncertainty in the choice of fits is the largest 
source of error in our determination of $m_N/m_\rho$.

\psfigure 3.0in -0.15in {FIG3} {fig3.ps} {$m_N/m_\rho$ as a function of
$am_\rho(m_q=0)$ with the ratio $m_\pi/m_\rho$ held fixed at its physical value.}

\section{QUENCHED SPECTRUM WITH IMPROVED STAGGERED QUARKS}

Up to now, most of the effort aimed at improving the quark action has
focused on Wilson quarks; however, some time ago Naik proposed 
adding third nearest neighbor hopping terms to the staggered quark 
action, adjusted to remove the order $a^2$ errors in the lattice quark 
propagator\cite{NAIK}. We are experimenting with
a tadpole improved version of this action to carry out a
quenched spectrum calculation with gauge configurations generated
using a tadpole-Symanzik improved gauge action\cite{PUREG}. We have generated
configurations at three different couplings, $\beta_{pl}=6.8$, 7.1
and 7.4, on lattices ranging in
size from $8^3\times 16$ to $16^3\times 32$. We have measured
the spectrum on these lattices with both the tadpole improved
Naik action and the standard Kogut-Susskind action. In Fig.~4
we show the Edinburgh plot for the Naik (crosses) and standard staggered
(diamonds) quark actions on $16^3\times 32$ lattices at $\beta_{pl}=7.4$. For
comparison, we also show quenched spectrum results with the standard gauge and
staggered fermion actions (fancy plusses) at $6/g^2=5.54$. This coupling
corresponds to approximately the same lattice spacing, as measured
by the rho mass, as the improved action at $\beta_{pl}=7.4$.
One sees from Fig.~4 that the improved gluon
action does lead to lower values of the nucleon to rho mass ratio
at small values of $m_\pi/m_\rho$, than the standard gluonic action. 
However, the Naik term has little effect at the values of the coupling
constants and quark masses that we have investigated to date.

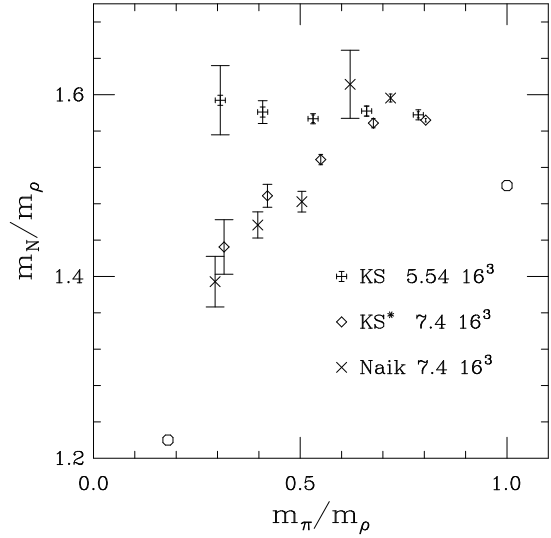
\psfigure 3.0in -0.15in {FIG4} {fig4.ps} {The Edinburgh plot for the
quenched spectrum with improved gluon action at $\beta_{pl}=7.4$.} 

An important test of improved staggered quark actions is the extent
to which they restore flavor symmetry. Our preliminary results indicate
that the mass difference between the $\pi$ and $\pi_2$ does decrease
with the improved gluon action, but that the Naik term has little impact. 
Although our tests of the tadpole improved Naik action have not been very
encouraging at the parameters studied to date, it should be noted
that this action has the advantage of significantly hastening the rate
at which thermodynamic quantities approach the continuum limit for
free quarks. One might therefore hope that it would lead to a significant
improvement in high temperature QCD calculations, and the Bielefeld
group has shown that this is the case\cite{BIELEFELD}. It would 
therefore seem that this action should be studied in greater detail. 

\section{SPECTRUM WITH TWO FLAVORS OF STAGGERED QUARKS AT $\bf 6/g^2=5.50$}

As part of our study of the decay constants of heavy-light mesons,
we are generating a set of lattices with two flavors of Kogut-Susskind
quarks at gauge coupling $6/g^2=5.50$. We are using quark masses
$m_q=0.10$, 0.05, 0.025 and 0.0125. All lattices have a time
dimension $N_t=64$. We use spatial volumes $20^3$ for the two
lighter masses, and $24^3$ for the two heavier ones. (The use
of larger spatial volumes for the heavier quark masses was
dictated by the peculiarities of the CM5 on which the heavier
quark mass calculations were performed).
We have calculated the spectrum using the first
1,000 equilibrated molecular dynamics time units for each quark
mass. The Edinburgh plot is shown in Fig.~5.
We have included quenched data for gauge couplings $6/g^2=5.70$,
5.85 and 6.15 for comparison. The data indicates a somewhat
faster fall off of $m_N/m_\rho$ than the quenched data at
$6/g^2=5.70$, which has a comparable lattice spacing. This is
to be expected because lighter dynamical quarks will
lead to a greater renormalization of the coupling.

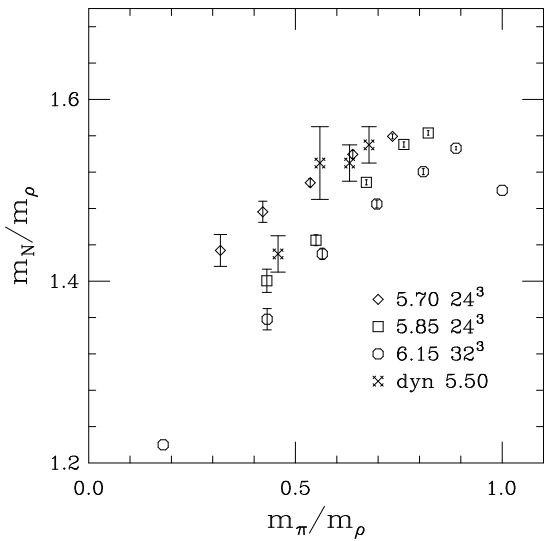
\psfigure 3.0in -0.20in {FIG5} {fig5.ps} {The Edinburgh plot
for full QCD with two flavors of staggered quarks at $6/g^2=5.50$.}

In Fig.~6 we plot the masses of the $\pi$ and the $\pi_2$ as a
function of $am_\rho$ to illustrate the extent of flavor symmetry
breaking in this data set. We also include very preliminary results
from a set of runs at $6/g^2=5.60$ on $24^3\times 64$ lattices
with quark masses 0.01, 0.02, 0.04 and 0.08. These results are
from approximately 600 equilibrated trajectories at each mass.

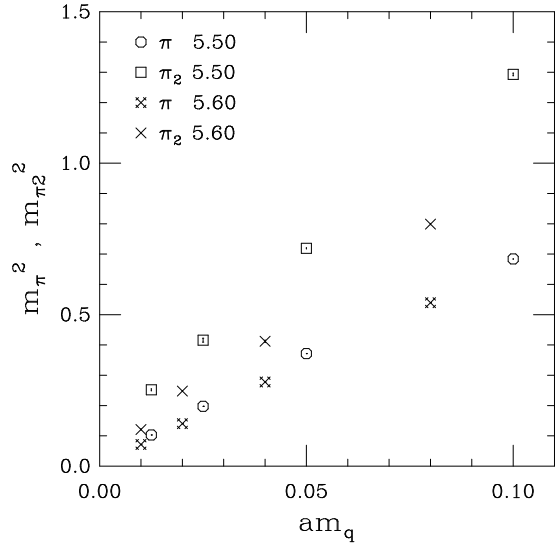
\psfigure 3.0in -0.15in {FIG6} {fig6.ps} {$m_\pi$ and $m_{\pi_2}$
as a function of $am_\rho$ for full QCD at $6/g^2=5.50$ and $5.60$.
The 5.60 results are quite preliminary.}

%\section*{Acknowledgements}
The calculations described above were carried out on Intel 
Paragons at Indiana University,
ORNL and SDSC, the TMC CM5 at NCSA,
the Cray T3D at PSC, and DEC Alphas at Indiana University and PSC.
This research was supported by the United States NSF and DOE.

\end{document}